\documentstyle[aps,multicol]{revtex}

\renewcommand{\narrowtext}{\begin{multicols}{2} \global\columnwidth20.5pc} 
\renewcommand{\widetext}{\end{multicols} \global\columnwidth42.5pc} 
\catcode`\@=11 
\def\slash{\@ifnextchar[{\@slash}{\@slash[\z@]}} 
\def\@slash[#1]#2{\setbox\z@\hbox{$#2$}\@tempdima\wd\z@\box\z@% 
\@tempdimb#1 \advance\@tempdimb-\@tempdima \kern\@tempdimb 
\hbox to\@tempdima{\hss\@makeslash\hss}} 
\def\@makeslash{$/$}                    % This is what is overstruck 
\catcode`\@=12 
\def\labl#1{\label{#1} 
} 
\def\beq#1{\begin{equation} \labl{#1}} 
\def\eeq{\end{equation}} 
 
\def\ep{\epsilon}

\def\ber{\begin{eqnarray}} 
\def\eer{\end{eqnarray}}

\begin{document} 
%%%%%%%%%%%%%%%%%%%%%%%%%%%%%%%%%%%%%%%%%%%%%%%%%%%%%%%%%%%%%%%%%%%%%%%%% 
%%%%%%%%%%%%%%%%%%%%%%%%%%%%%%%%%%%%%%%%%%%%%%%%%%%%%%%%%%%%%%%%%%%%%%%% 
%%%%%%%%%%%%%%%%%%%%%%%%%%%%%%%%%%%%%%%%%%%%%%%%%%%%%%%%%%%%%%%%%%%% 
\title{Tunnelling Spectroscopy of Localized States near  
the Quantum Hall Edge} 
\author{A. Alekseev,  V. Cheianov}  
\address{\it Uppsala University, Uppsala \\ S-751 08,  Sweden  
\\ E-mail:sheianov@helene.teorfys.uu.se }  
\author{ A. P. Dmitriev,  V. Yu. Kachorovskii}  
\address{ A.F.Ioffe Physical-Technical Institute,  
St.Petersburg, \\ 194021, Russia \\ E-mail:  
dmitriev@vip1.ioffe.rssi.ru }  
\date{April, 1999} 
\maketitle 
%%%%%%%%%%%%%%%%%%%%%%%%%%%%%%%%%%%%%%%%%%%%%%%%%%%%%%%%%%%%%%%%%%%%%%% 
{\tightenlines 
\begin{abstract} 
In the paper we discuss experimental results of  
M. Grayson {\it et al.}  on tunneling $I$-$V$  
characteristics of the quantum Hall edge.  
We suggest a two step tunneling mechanism involving  
localized electron states near the edge, which might  
account for discrepancy between the experimental data 
and the predictions of the chiral Luttinger liquid theory 
of the quantum Hall edge.  
\vskip 0.1cm 
 
 
\end{abstract} 
} 
\narrowtext 
 
Measuring the tunnelling current from a normal metal to 
a quantum Hall edge is an attractive way to observe the Luttinger 
liquid behaviour of QH edge states. 
Scaling invariance of the Luttinger liquid should leave a clear 
signature in the $I$-$V$ characteristic of the tunnelling current 
in the form of a power law dependance of the current on the 
applied voltage \cite{kane} 
\begin{equation} 
I \sim 
V^{\alpha}. 
\label{i} 
\end{equation} 
Moreover, 
the value of the tunnelling exponent $\alpha$ would give the 
knowledge of the decay law of the electron Green's function 
at the edge which would provide a direct check of the 
chiral Luttinger liquid theories of the edge states \cite{kane,wen,FrK}. 
 
The predictions of the chiral Luttinger liquid theory 
of the QH edge states can be summarized as follows.

(1) If a QH state belongs to 
the principal (Laughlin) sequence of 
filling factors $\nu =1/(2p+1)$ ($n$ integer) it 
supports one gapless chiral edge mode. In this 
case the tunnelling exponent is given by 
$$\alpha= 1/ \nu. $$ 
Both at very low and very high bias there can be deviations from the 
simple power law (\ref{i}) even within the Luttinger liquid picture \cite{frad}. 
The deviation at low bias is a finite temperature effect, whereas 
at high biases the deviations 
depend on the way the tunnelling contact 
is arranged. 
By a proper 
choice of experimental conditions one can, create a broad 
window in voltage, where the power law 
dependance (\ref{i}) holds.

(2) In a more general case of incompressible QH states corresponding to Jain filling 
factors $\nu= N/(2Np+1)$ 
there are $|N|$ edge modes and the calculations 
\cite{kane,wen} based on the Luttinger liquid picture predict that 
\begin{equation} 
\alpha = 1 + |2p + 1/N| - 1/|N|. 
\label{al2} 
\end{equation} 
In \cite{halp} 
an alternative theory was developed capable of 
treating compressible states near the filling $1/2$. The 
expression for $\alpha(\nu)$ was obtained which in the limit of 
vanishing compressibility 
coincides with (\ref{al2}) for Jain filling factors. This theory 
predicts that 
the slope of the tunnelling exponent $d\alpha/ d\nu $ behaves 
discontinuously as a function of 
$\nu$ at points $\nu=1$, $\nu=1/2$. 
 
Recent tunnelling experiments \cite{chang} gave the  
following results: 
 
(1) At low temperatures the $I$-$V$ characteristic exhibits 
the power-law behaviour (\ref{i}) up to 6 decades in current.
The power-law $I$-$V$ characteristics are observed independently
of whether the 2DEG is in a compressible or in an incompressible
state.

(2) The tunnelling exponent $\alpha$ varies continuously with the
filling factor $\nu$. The dependance of the
tunnelling exponent on the filling factor is well approximated
by the linear law $\alpha=1/\nu$.

One can see that the second item is in an obvious
contradiction to the predictions of the chiral Luttinger
liquid theory.

This disagreement puts in doubt the generally accepted
theories of the fractional
quantum Hall edge and needs to be explained either within these
theories or by developing a new theoretical approach. That is why
a lot of attention has been paid to the problem of late
\cite{conti,Han,LW,Frad,ZM,Kh}.

In the discussion below we restrict ourselves to the
incompressible case only. In this case the quantum Hall edge
is believed to be described by a chiral Luttinger liquid theory.
In it's standard form this theory \cite{kane,wen} claims
that there are several edge chiral Luttinger modes which can be
separated into one charged mode and a number of neutral ones.
These modes are described by bosonic fields $\varphi_0,
\varphi_1, \dots ,\varphi_{N}$ 
propagating along
the edge and having the dispersion
\begin{equation} \label{disp}
\omega_j(q)=s_j q.
\end{equation}
The anomalous exponent
$\alpha$ in the $I$-$V$ characteristic (\ref{i}) is simply related
to the asymptotic behaviour of the single particle Green's
function of an electron in the QH system \cite{mahan}:
\begin{equation}
G(t)=-i\langle T \psi(x,t) \psi^\dagger(x,0) \rangle \sim \frac{1}{t^\alpha}
\label{Green}
\end{equation}
In the chiral Luttinger liquid theory
(\ref{Green}) is a correlation function of the electron operators
\begin{equation}
\psi=\exp {i(\varphi + \sum_{j=1}^{N-1}m_j \varphi_j)}
\label{opr}
\end{equation}
with the critical exponent
\begin{equation}
\alpha = \frac{1}{\nu} + 2\sum_{j=1}^{N-1} m_j^2 \delta_j
\label{Alp}
\end{equation}
where $\delta_{j}$ are the dimensions of the operators
$\exp(i \varphi_j)$. From (\ref{Alp}) we see that the critical
tunnelling exponent cannot be smaller than $1/\nu$. What is more,
for all $1/3 <\nu <1$ it turns out to be bigger than $1/\nu$
and given by (\ref{al2})
because $m_j$ in the expansion (\ref{Alp}) must be
nonzero due to the requirement that the operator (\ref{opr})
be fermionic \cite{wen,FrK} .

An attempt to solve the contradiction
between the theory and the experiment was made in \cite{LW,Frad}.
The main idea of these works is that the experimentally
observed tunnelling exponent is consistent with the chiral Luttinger
liquid picture under the assumption that the neutral modes
are non-propagating (or their propagation velocities are
negligible). In our opinion the weakness of this approach is
that it's central assumption has no sufficient physical justification.
In the experiment,
the power law $I$-$V$ characteristic is observed in a broad
(about two decades) voltage range. For the picture \cite{LW,Frad}
it implies that the velocities of the neutral edge modes should
differ from the velocity of the charged mode by minimum two orders of
magnitude. This difference is attributed
to the Coulomb interaction, whose contribution to the velocity of the
charged mode is evidently
larger than to the neutral ones. However the edge
modes must have some transverse structure on the scale 
of the magnetic length. As a result,
the neutral modes carry transverse dipole (and higher order) moments
giving rise to the multipole interaction
contributing to their velocity. Simple analysis shows that (see \cite{ZM}
and considerations below)
\begin{equation}\label{ln}
\frac{s_0}{s_j} \sim \ln\left(\frac{1}{qa}\right)
\end{equation}
where $a$ is the quantum well width. The logarithmic factor on the
r.h.s of (\ref{ln}) evidently cannot account for the two decade difference
needed for the picture of \cite{LW,Frad}.

The first term on the r.h.s of Eq. (\ref{Alp}) corresponds to the shakeup
of the charge relaxation mode at the edge. If one neglects the
contribution of other bosonic modes the experimentally observed
tunnelling exponent will be regained.
This is exactly the way
the problem is treated in the framework of the
independent boson model (IBM) \cite{mahan,conti,Kh}, where a
single localized electron electrostatically interacts with the
hydrodynamic charged edge mode in the incompressible case \cite{conti}
or with the bulk charge relaxation modes
in the compressible case \cite{Kh}. Although the IBM gives a
correct tunnelling exponent it says nothing about the
physical origin of the localized states and it's relevance to the
experiments is not clear unless the nature of these states is
specified.
In particular, understanding what these states are is important
since the results which can be obtained in the framework
of the IBM are very sensitive to the choice of the energy position
of the localized state.

In our opinion, a good agreement between the observed universality of the
$I$-$V$ characteristics and the IBM description indicates
that near the edge of the QH liquid there exist some low energy
electronic states other than the excitations of the chiral Luttinger
liquid. Electrons tunnelling from the metal into these states
electrostatically interact with the charged mode of the
edge collective excitations. Below we suggest a
model of the edge where the tunnelling current is transmitted in
a two-step process which involves localized states in the bulk at the
intermediate stage. We show that this processes gives an experimentally
observed $I$-$V$ exponent providing that the intermediate localized states
are spatially separated from the edge and their energy distribution
function decays exponentially in the gap of the incompressible
states like in the integer quantum Hall regime \cite{Wegner}.

The series of well established plateaus observed
in the experiment \cite{chang} indicates that in the gaps of
incompressible QH states there exists a finite density of bulk
localized states
$ g(\ep) $
created by the random impurity potential.
An electron may tunnel into the QH edge in a two step process:
first it tunnels into a localized state, where it may stay
for some time $t^*$, and then decays into the the edge mode
due to a finite hybridization between the edge and bulk states.
If the voltage $V$ satisfies the condition
$$\frac{\hbar}{eV}\ll t^*$$
then the second step of the tunnelling process does not affect the
$I$-$V$ characteristics.

The two step tunnelling mechanism competes with the direct tunnelling
into the edge states. There are two reasons why the two step mechanism
may be more efficient:

(1)  The tunnelling exponent associated with this mechanism turns out
to be smaller than the one given by
Eq. (\ref{al2}) (see below).

(2) Unlike the tunnelling into the edge state the
tunnelling into the localized states does not require any additional
scattering off the impurities in the metal \cite{chang} (in the absence of
impurities in the metal direct tunnelling into the edge state would
have been altogether impossible due to the momentum conservation).

On the time scale $t^*$ the tunnelling process is described by the
IBM model. If the QH system is incompressible,
the only
contribution to the polarization of the QH medium comes from
gapless edge modes. An electron can polarize both charged and
neutral modes (because they carry multipole moments).

First we consider the contribution of the charged mode.
The corresponding Hamiltonian reads
\begin{equation}
H = \sum_n (\epsilon_n- \hat w_n)a_n^\dagger a_n
%\int dx \hat \rho(x) U_n(x)
+ \frac{1}{2}\int dx dx' \hat \rho(x) v(x-x') \hat \rho(x').
\label{ham}
\end{equation}
Here $a_n$ is the annihilation operator of an electron in
the localized state with the energy $\epsilon_n$ and the wave function
$\psi_n({\bf r})$, $e \hat \rho(x)$ is the charge density operator
of the edge plasmon,
$v(r)=e^2/\kappa r$ is the Coulomb potential with the
dielectric constant $\kappa$. The term

\begin{equation}
\hat w_n=\int dx \hat \rho(x) U_n(x)
\label{wn}
\end{equation}
stands for the
electrostatic interaction between the edge plasmon and the
electron in the localized state. Here
\begin{equation} \label{Un}
U_n(x)=\int d^2 {\bf r}' v(|{\bf r-r'}|) |\psi_n({\bf r'})|^2
\end{equation}
is the potential induced by the localized state at the edge ${\bf r }=(x,0)$.
The charge density operator $e \hat \rho$ of the
edge plasmon is given by
$$ \rho = \sum_{q>0} i \sqrt{\frac{\nu q}{2\pi L}}(
b_q e^{iqx}-b_q ^\dagger e^{-iqx}),$$
where $L$ is the length of the edge.

The Hamiltonian (\ref{ham}) is diagonalized by the canonical
transformation to the new fermionic operators $\bar a_n$ related to
the operators in \
\cite{mahan}
\begin{equation}\label{Sol}
a_n
= \bar a_n e^{- i\Phi_n}  = \bar a_n T \exp \left(- i \int_{-\infty}^t dt' \hat w_n(t')\right)
\end{equation}
where $\rho(x,t)$ is the charge density operator in the
interaction representation, $T$ is the time ordering operator.
Introducing the field $\phi$, such that $\rho=\nu/2\pi\partial_x \phi$ and
taking into account that in the interaction representation it's dynamics is
given by (\ref{disp}) we find that the operator $\Phi$ in
(\ref{Sol}) is given by
\begin{equation}
\Phi_n(t) = \frac{\nu}{2 \pi s_0 } \int dx U_n(x)\hat\phi(x,t),
\end{equation}
where the velocity of the charged mode reads
\begin{equation}
s_0 = \frac{e^2 }{\kappa \pi \hbar} \nu \ln\left(\frac{1}{qa}\right)
\label{s0}
\end{equation}
Green's function (\ref{Green}) of an electron in the $n$-th localized state
is given by
\begin{equation}
G_n(t)=-i e^{i\tilde \epsilon_n t} (1-n_F(\tilde \epsilon))
\langle T e^{-i \Phi_n(t)}  e^{i \Phi_n(0)}\rangle  \\ ,
\end{equation}
where $\tilde \epsilon_n$ is the exact energy of the localized eigenstate
dressed by the charge mode relaxation.

The factor
\begin{equation}
 \langle T e^{-i \Phi_n(t)}  e^{i \Phi_n(0)}\rangle = \langle T e^{-i \int_0^{t'} \hat w(t')}\rangle
\label{T}
\end{equation}
 is
responsible for the suppression of the tunnelling density of
states due to interaction of the electron with the charged mode.
At large times, the main contribution to the
this factor comes from the long-wavelength limit. In our case this limit
is defined by $qd_n \ll 1 $, where $d_n $ is the distance from the edge to
the localized state (below it will be argued that $d_n>l$).
The asymptotic form of this factor does not depend on $n$ and is
given by
$$\langle T e^{-i \Phi_n(t)}  e^{i \Phi_n(0)}\rangle \sim \frac{1}{t^{1/\nu}}.$$

In the standard approximation \cite{mahan} where the dependence of the tunnelling
matrix element on the energy is neglected we obtain
 $$
I(V) \sim \int_0^{eV} d\epsilon g(\mu + \epsilon) (eV
-\epsilon)^{1/\nu},
$$
here $\mu$  is the chemical potential of the QH system and
$g(\epsilon)$ is the density of the localized states.

%It is clearly seen from Eq.(\ref{current3}) that
The tunnelling exponent $\alpha$ is determined by the behaviour of the
%the energy scale $\Gamma$ of
%the decay of the
density of localized states in the vicinity of the
Fermi energy. As far as we consider the incompressible QH
liquid, the Fermi level must lie in the gap of the volume
excitations.  It looks very natural to assume that the density of localized states
in the gap decays rapidly with the energy (just as in the case
of the integer QH effect). If the energy scale $\Gamma$ of the
decay is smaller than $eV$ the tunnelling current is given by
$$ I(V) \sim \Gamma V^{1/\nu} $$ and $\alpha =1/\nu
$. In the opposite limiting case ($\Gamma > eV $) the tunnelling
current is given by
$ I\sim V^{1/\nu+1} $.

Next we discuss the interaction of a tunnelling electron
with the multipole moments of the neutral modes.
This has an analogy with the
model of smooth edge considered in \cite{conti0}, where the tunnelling
exponent is much larger than $1/\nu$ due to the contribution
of the modes carrying multipole moments.

Most important is the interaction with
the mode responsible for the
dipole moment. In contrast to the case of the charged mode this
interaction depends on the concrete model of the edge, esp. on
the transversal structure of the neutral modes.
We take this interaction into account phenomenologically.
The model Hamiltonian describing the interaction of localized
states with the neutral modes
should have the same form as
(\ref{ham}) with two differences. First, one should replace
 $v(|{\bf r-r'}|)$ in
Eq. (\ref{Un}) by the dipole interaction
$\delta y \partial_x  v(|{\bf r-r'}|)  $ , where
the quantity $\delta y$  is of  the order of the
width of the edge strip and depends of the concrete model of the edge. In what follows
we assume  $\delta y  \approx l$.
Second difference is that in contrast with (\ref{s0})
the velocity of the dipole mode
does not contain the Coulomb logarithm $s_1 \sim \frac{e^2 }{\kappa \pi \hbar} \nu $.
Taking into account the dipole interaction in the framework of IBM shows that
the
 factor (\ref{T}) falls off more rapidly at large $t$:
$$\:\: \langle T \exp(-i \int_0^{t'} \hat w(t'))\rangle\sim t^{-[\frac{1}{\nu} +
\frac{l^2}{d_n^2} \frac{1}{\nu}]} .$$
 For the
tunnelling exponent at $\Gamma < eV $ we get

\begin{equation}
\alpha(d_n)= \frac{1}{\nu} +
\frac{l^2}{d_n^2} \frac {1}{\nu}.
\label{alpha3}
\end{equation}
The tunnelling probability falls off rapidly with the tunnelling
distance as $$ \exp{(-d_n^2/l^2)} $$ leading to
decrease of tunnelling current.
 On
the other hand, the larger is $d_n$   the smaller is $\alpha(d_n)$.
This leads to $I$ increasing
with $d_n$ at small values of the applied voltage.
 As a result,
there exist optimal values of the tunnelling distance $d_{\rm opt} > l$ and of the tunnelling exponent $\alpha_{\rm opt}$.
The values of $d_{\rm opt} $ and of 
 $\alpha_{\rm opt}$ can be determined using the following
estimate of
the tunnelling current related to the $n$-th localized state
$$I_n(V) \sim e^{-\frac{d_n^2}{l^2}} \left(\frac{eV}{E_0}\right)
^{\alpha(d_n)}. $$
Here $E_0$ is large energy ($E_0 \ll eV$) , which is of the order
of the Fermi energy. From this equation we easily get
$$\alpha_{\rm opt}= \frac{1}{\nu}\left(1 +  \left(\frac{\nu
}{\ln(\frac{E_0}{eV})}\right)^{1/2}\right) . $$
It can be seen, that at small voltages
%the optimal tunnelling
%distance is larger than $l$, the electron interacts effectively
%with the charged mode only and
the optimal tunnelling exponent
is  close to the experimentally observed value

$$\alpha_{\rm opt} \approx \frac{1}{\nu} . $$

We would like  to emphasise that this result is related to the fact
that the electron tunnels at the distance larger than the width
of the edge strip $l$. On the contrary, in the model used in
Ref. \cite{conti0} it was implicitly assumed that $d_n \sim
\delta y $ and the
tunnelling exponent was found to be much larger than $ 1/\nu$.

We conclude that the tunnelling process as a whole looks as
follows. First the electron tunnels in the localized state
and during the time $d_{\rm opt}/s_0$ polarizes the
charged mode. As a result the positive screening charge $e$ is attracted
to the edge in the region of length $d_{\rm opt}$ and the
compensating negative charge $-e$ is carried away by the the charged
mode with the velocity $s_0$. On time scale $t$ such that 
$ d_{\rm opt}/s_0 <t< t^*$ there exists a dipole formed by the 
localized electron and the
screening positive charge. After the time $t^*$
this dipole vanishes due to the tunnelling of the electron from
the localized state into
the edge.

\subsection*{Acknowledgements}

The authors would like to thank J. Fr\"ohlich for
stimulating discussions.
V.C. is grateful to B. I. Halperin for important remarks
and for hospitality at the Lyman Lab. 
We hould like to thank L. Levitov and A. Chang for valuable 
comments and information about their work.

\widetext

%******************************************************************* 
\end{document}